\documentclass[aps,prd,superscriptaddress,showkeys,showpacs,twocolumn,a4paper]{revtex4-1}
\usepackage{ulem}
\usepackage{color}
\usepackage{graphicx}
\usepackage{appendix}
\usepackage{epsfig}
\usepackage[colorlinks=true,linkcolor= teal, citecolor = blue, urlcolor = blue ]{hyperref}
\usepackage[T1]{fontenc}

\usepackage{epstopdf}
\usepackage{amsmath}
\usepackage{subfig}
\usepackage{comment}
\usepackage{float}
\usepackage{float}
\usepackage[dvipsnames]{xcolor}
\newcommand{\be}{\begin{equation}}
\newcommand{\ee}{\end{equation}}
\newcommand{\ba}{\begin{eqnarray}}
\newcommand{\ea}{\end{eqnarray}}
\newcommand{\nn}{\nonumber\\}
\UseRawInputEncoding

\begin{document}

\title{Subdiffusion of  heavy quark in hot QCD matter by the fractional Langevin equation}

\author{Jai Prakash}
\email{jaiprakashaggrawal2@gmail.com}
\affiliation{Department of Physics, Indian Institute of Technology Bombay, Mumbai 400076, India}

\affiliation{Department of Theoretical Physics, Tata Institute of Fundamental Research, \\
Homi Bhabha Road, Mumbai 400005, India}



\begin{abstract}

 The subdiffusion phenomena are studied for heavy quarks dynamics in the hot QCD matter. My approach aims to provide a more realistic description of heavy quark dynamics through detailed theoretical analyses and numerical simulations, utilizing the fractional Langevin equation framework with the Caputo fractional derivative. I present numerical schemes for the fractional Langevin equation for subdiffusion and calculate the time evolution mean squared displacement and mean squared momentum of the heavy quarks. My results indicate that the mean squared displacements of the heavy quarks for the subdiffusion process deviate from a linear relationship with time. Further, I calculate the normalized momentum correlation function, kinetic energy, and momentum spread. Finally, I show the effect of subdiffusion on experimental observables, the nuclear modification factor. 
\end{abstract}



\maketitle

 \section{Introduction}
 \label{intro}

Ultra-relativistic collisions of heavy ions at the Relativistic Heavy-Ion Collider (RHIC) and the Large Hadron Collider (LHC) have proposed the presence of quark-gluon plasma (QGP)  \cite{STAR:2006vcp, Adams:2005dq, PHENIX:2004vcz, ALICE:2010khr, BRAHMS:2005gow}. This QGP phase exists fleetingly, lasting only a few fm/c (4$-$5 fm/c at RHIC and 10$-$12 fm/c at LHC) ~\cite{vanHees:2004gq, Rapp:2009my}. Conversely, exploring the properties of QGP through the heavy quarks (charm quark and bottom quark) dynamics is a topic of high contemporary interest \cite{Song:2015sfa, Andronic:2015wma, Dong:2019unq,  Aarts:2016hap, Uphoff:2011ad, Plumari:2017ntm, Gossiaux:2008jv, Prakash:2021lwt, Prakash:2023wbs, Prakash:2023hfj, Singh:2023smw, Kurian:2020orp, Cao:2016gvr, Mazumder:2011nj, Zhang:2022fum, PhysRevD.103.054030, Jamal:2021btg,Jamal:2020emj, Sun:2023adv,Plumari:2019hzp, Prakash_2024, Du:2023ewh,Shaikh:2021lka,Kumar:2021goi,das2022dynamics,PhysRevLett.132.222301,Altmann:2024kwx,PhysRevLett.130.231902,ZACCONE2024116483,PhysRevD.109.L011503,PhysRevD.106.034032,Jamal_2024,PhysRevD.109.116008,Das:2024vac,Chandra:2024ron,K:2023dum,Pooja:2023gqt}. Heavy quarks (HQs) are almost formed in initial hard scatterings of very early phases of ultrarelativistic heavy-ion collisions (HICs) and conserved during the evolution. The HQs take a longer time to become thermalized than lighter medium particles. Due to their significantly larger masses compared to the typical temperatures achieved in HICs, their diffusion through the hot QCD matter can be modelled as a Brownian particle. Slowly moving HQs undergo a Brownian motion in a medium of light quarks and gluons \cite{Moore:2004tg}. The HQs are valuable probes for studying the time evolution of the QGP medium. Their dynamics can be studied using the Langevin equation, where the interaction of the HQs with the medium is taken through the transport coefficients \cite{Cao:2018ews,Das:2013kea}. The standard Langevin equation (LE) is used to study Brownian motion, assuming normal diffusion. This means that the mean-squared displacement ($\langle x(t)^2\rangle$) of the particle increases linearly with time when observed over a long period \cite{kubo2012statistical}. I use the LE to study the  HQs dynamics in hot QCD matter  \cite{PhysRevC.86.034905, PhysRevC.93.014901, PhysRevC.84.064902,vanHees:2007me, Das:2013kea}. 

{ In conventional Langevin approaches, the description of the HQs energy loss is twofold: while these approaches are effective in modelling elastic energy loss, they face significant challenges in accurately capturing radiative energy loss.  It is possible to address this by introducing an additional force term that accounts for emitted gluon radiation. This method has been effectively employed in previous studies \cite{PhysRevC.88.044907}, demonstrating that with appropriate modifications, conventional Langevin frameworks can also handle radiative energy loss scenarios. Further, this approach describes the elliptic flow ($v_2$) of heavy hadrons \cite{PhysRevC.97.014907}. 
{$R_{AA}$ and the $v_2$ are primary observables associated with HQs measured at both RHIC 
and LHC energies. However, the first experimental results were a surprise. Nearly
all existing models encounter challenges in effectively describing both the $R_{AA}$ and $v_2$
simultaneously \cite{Das:2015ana,SUN2019134933}. This challenge is not limited to RHIC energy; similar difficulties
are also evident in results from LHC energy collisions. Toward the understanding of the $R_{AA}$ and $v_2$
puzzle for the HQs, one can consider the subdiffusion for the HQs motion in the QGP medium.} This raises the question: Could the anomalous process for the HQs provide a better explanation for the observed experimental data? To explore this, I need to study if the anomalous behaviour seen in nonrelativistic Brownian particles also occurs in relativistic ones. By doing so, I can see if anomalous motion might explain the discrepancies in energy loss for HQs in the QGP medium. 
In recent years, there has been increasing interest in studying processes in random disordered materials and systems that show chaotic behavior. In these systems, the duration of any regular motion is unpredictable. A well-known example is anomalous diffusion. Anomalous diffusion has applications in many fields, such as nonlinear dynamics \cite{ZASLAVSKY2002461}, in complex systems \cite{doi:10.1080/08927020008044119}, molecular chemistry \cite{ZUMOFEN1994303}, biology \cite{PhysRevLett.95.260603}, and the finance market \cite{PhysRevE.65.037106,PhysRevE.66.046118}.  In anomalous diffusion, the $\langle x(t)^2\rangle$ of a Brownian particle does not follow Einstein's law. Instead, it is described by a power function with a non-integer exponent as \cite{balescu1995anomalous,BOUCHAUD1990127, PhysRevLett.85.5655, PhysRevLett.104.238102,oliveira2019anomalous,sokolov2005diffusion,lutz2001fractional, PhysRevE.107.024105,PhysRevLett.125.240606,PhysRevE.104.024115,PhysRevE.108.034113} : 
\begin{align}\label{MSD}
\langle  x(t)^2\rangle \propto t^{\nu}, \  {\nu} \neq 1, 
\end{align}
in this equation, $t$ is the evolution time of the Brownian particle; normal diffusion occurs when ${\nu} = 1 $. Subdiffusion happens when ${\nu}< 1 $ \cite{Mainardi2007SubdiffusionEO}, and superdiffusion occurs for ${\nu}> 1 $ \cite{METZLER20001}. The main goal of this article is to consider the subdiffusion for the  HQs dynamics of the QGP, the effect of the superdiffusion is already discussed in \cite{PhysRevD.109.114004}. This anomalous diffusion has been observed in many physical systems \cite{PhysRevLett.99.200601,PhysRevLett.93.180603}.
To study anomalous diffusion, researchers have proposed fractional Langevin equations (FLE) \cite{coffey2012langevin,MEGIAS2024138370}. Mainardi and his group made significant contributions by introducing an FLE in their pioneering work \cite{MR1611585, mainardi2008fractional, MR1611587}. The Langevin method has predominantly utilized the Caputo fractional derivative for studying anomalous diffusion \cite{guo2013numerics,li2013finite,miller1993introduction}. However, there are various formulations for fractional derivatives, including the Riemann-Liouville derivative \cite{carpinteri2014fractals}, Riesz derivative \cite{Agrawal_2007}, Feller derivative \cite{Feller1971}, and others. Another effective approach for studying anomalous diffusion involves examining the fractional diffusion equation, the fractional Fokker-Planck equation \cite{PhysRevLett.82.3563}, and the Chapman-Kolmogorov equation \cite{PhysRevE.62.6233}.

Anomalous diffusion has been studied in the hot QCD matter in Refs. \cite{Caucal:2021lgf,Csanad:2007fr,Bhattacharyya:2024hku} and also the dynamics in fractal spaces in QGP \cite{MEGIAS2024138370}. With these motivations, I aim to study the subdiffusion for the dynamics of the HQ in hot QCD matter; I use the FLE, which is a more generalized form of the traditional LE. Unlike the LE, the FLE uses fractional-order derivatives. In our analysis, I exclusively use the Caputo fractional derivative to study the anomalous diffusion of the HQa through the FLE~\cite{10.1111/j.1365-246X.1967.tb02303.x}. This work might be the first to introduce subdiffusion for the  HQ dynamics within the QGP medium.
I anticipate a significant outcome in our study: the presence of subdiffusion will lead to a change in energy loss experienced by the HQs within the QGP medium. This conclusion is supported by studying various key parameters, including the  ($\langle x(t)^2\rangle$), mean-squared momentum ($\langle p(t)^2\rangle$), kinetic energy, normalized momentum correlation function, the momentum distribution ($dN/dp_T$), and nuclear modification factor ($R_{AA}$) of the HQs. 

The article is structured as follows: Section II introduces the formalism, provides an analytical solution to the FLE for non-relativistic heavy particles using the Laplace transform technique, and presents a numerical scheme for solving the relativistic FLE. Section III presents the results. Finally, Section IV offers a comprehensive summary of our conclusions.

\section{Formalism}

\label{NONRELATIVISTIC LIMIT}
For illustrative purposes, I consider a scenario where a Brownian particle with mass, $M$, undergoes one-dimensional (1D) motion in the non-relativistic limit. I use the FLE to describe the dynamics of this Brownian particle in the medium. In this framework, the evolution of the particle's position and momentum is characterized by fractional derivatives of orders $\beta$ and $\alpha$ \cite{kobelev2000fractional,li2012,lim2009modeling,PhysRevD.109.114004},
\begin{align}\label{Langevin_x}
&^{C} D^\beta_{0+}x(t)=\frac{p(t)}{M},
\\
&^{C} D^\alpha_{0+}p(t) = -\gamma p(t)+\xi(t),
\label{Langevin_p}
\end{align}
in this framework, the momentum of a particle at a time, $t$, is denoted as $p(t)$, and its position as $x(t)$. The Caputo fractional derivative is represented as $^{C}D_{0+}^{\alpha}$ and $^{C}D_{0+}^{\beta}$, with $\alpha$ and $\beta$ being the fractional parameters respectively, where $n-1 < \alpha \leq n$ and $n-1 < \beta \leq n$ ($n$ is a natural number). 
A Brownian particle encounters two distinct forces: the dissipative force, characterized by the drag coefficient ($\gamma$), and the stochastic force, denoted as $\xi(t)$. The stochastic force describes random noise ($\eta$), commonly referred to as white Gaussian noise. This is characterized by the instantaneous decay of correlations of the noise, often described as $\delta$  correlation.
The random force $\xi(t)$ satisfies certain properties, such as:

\begin{align}\label{corr}
&\langle\xi(t)\xi(t')\rangle=2\mathcal{D}\delta(t-t'), \\
 &\langle\xi(t)\rangle=0,
\end{align}

{ where $\mathcal{D}$ is the diffusion coefficient in momentum space}. The drag coefficient, $\gamma$, is related to the $\mathcal{D}$ through the Fluctuation-Dissipation Theorem (FDT) at temperature ($T$) as follows:
\begin{align}\label{FDT}
\gamma = \frac{\mathcal{D}}{MT}.
\end{align}
In Eqs.~\eqref{Langevin_x},\eqref{Langevin_p} the Caputo fractional derivative, $^{C} D^\mu_{0+}u(t)$ \cite{10.1111/j.1365-246X.1967.tb02303.x} is written  as, 

\begin{align}\label{Caputo}
 ^{C} D^\mu_{0+}u(t) = \frac{1}{\Gamma({n-\mu})}\int_0^t \frac{u^{(n)}(s)}{(t-s)^{1+\mu-n}}ds,   
\end{align}
 where {$n =\lceil{Re(\mu)}\rceil+1$,} $u^{(n)}$ denotes the $n^{th}$ derivative of $u$, and $\Gamma(\cdot)$ denotes the gamma function. 
For $0<\mu\leq 1$, the fractional derivative corresponds to subdiffusion and is given by,
\begin{align}\label{subD}
 ^{C} D^\mu_{0+}u(t) = \frac{1}{\Gamma({1-\mu})}\int_0^t \frac{u^{(1)}(s)}{(t-s)^{\mu}}ds,    
\end{align}
where $u^{(1)}(s)$ denote the first derivative of $u(s)$. {To demonstrate the connection between the FLE and the standard Langevin equation, I can solve Eq. \eqref{subD} by setting the order of the derivative $\mu$ = 1; the fractional derivative reduces to a standard first-order derivative. Specifically, one can simplifies  $^CD^{\mu}_{0^+} u(t)$  to $ u^{(1)}(t) $ in this limit. As a result, the FLE (Eqs.~\eqref{Langevin_x}, \eqref{Langevin_p}) naturally reduces to the standard Langevin equation, which is governed by first-order derivatives only. }
In the subsequent subsection, I solve analytically the FLE of nonrelativistic Brownian particles.

\subsection{Analytical solution of FLE}

To solve the FLE (Eqs.\eqref{Langevin_x}, \eqref{Langevin_p}) for non-relativistic Brownian particle motion, I utilise the Laplace technique to derive analytical solutions for $\langle p^2(t) \rangle$ and $\langle x^2(t) \rangle$. The Laplace transformation of the Caputo derivative for $n-1 < \mu \leq n$ \cite{podlubnyacademic} is written as follows
\begin{align}\label{laplace}
 \mathcal{L}\left[^{C} D^\mu_{0+}g(t)\right](z)=z^\mu\widehat{g}(z)-\sum_{k=0}^{n-1}z^{\mu-k-1}\left[g^k(0)\right].  
\end{align}
To calculate the solution for $p(t)$ and $x(t)$ of the Brownian particle in a subdiffusive process, one can take the Laplace transform of Eqs. (\ref{Langevin_x}), (\ref{Langevin_p}) using Eq. \eqref{laplace}. For the subdiffusion, the fractional derivatives are $0 < \alpha \leq 1$ and $0 < \beta \leq 1$. For simplicity, I set $M = 1$ \cite{Sandev2012}, making the calculations clearer as,

\begin{align}\label{sublaplaceofx1}
 &z^\beta\widehat{x}(z)-z^{\beta-1}x(0) = \widehat{p}(z),\\  
\label{sublaplaceofv1}
&\widehat{v}(z)= \frac{z^{\alpha-1}}{z^\alpha + \gamma}p(0) + \frac{\widehat{\xi}(z)}{z^\alpha + \gamma},
\end{align}
 for the solution $x(t)$, substituting Eq. (\ref{sublaplaceofv1}) into Eq. \eqref{sublaplaceofx1},  I obtain 
\begin{align}\label{sublaplaceofx}
  \widehat{x}(z)=\frac{x(0)}{z}   +\frac{z^{\alpha-\beta-1}}{z^\alpha + \gamma}p(0) + \frac{z^{-\beta}}{z^\alpha + \gamma}\widehat{\xi}(z).
\end{align}
Now, by using the inverse Laplace transformation of Eqs. \eqref{sublaplaceofv1}, \eqref{sublaplaceofx} I obtain the analytical solution of $p(t)$ and $x(t)$ for the subdiffusion as
\begin{align}\label{sublaplaceofv}
 p(t) = &\;  E_{\alpha,1}(-\gamma t^\alpha)p(0)  \nn 
  &+ \int_0^t (t-s)^{\alpha-1}E_{\alpha,\alpha}(-\gamma(t-s)^\alpha)\xi(s)ds,\end{align}
  \begin{align}\label{sublaplaceofX}
  x(t)= &\;x(0)+ t^\beta  E_{\alpha,\beta+1}(-\gamma t^\alpha)p(0)   \nn & +\int_0^t (t-s)^{\beta+\alpha-1}E_{\alpha,\beta+\alpha}(-\gamma(t-s)^\alpha)\xi(s)ds.
  \end{align}
  The two-parameter Mittag-Leffler function is described as: 
  \begin{align*}
 E_{\alpha,\beta}(z) = \sum^{\infty}
_{j=0}\frac{z^j}{\Gamma(j\alpha + \beta)}, 
\end{align*}
for $\alpha>0$, $\beta>0$, and a complex number $z$ \cite{erdelyi1953bateman}. The Laplace transform of the two-parameter Mittag-Leffler function \cite{MR1890104} is,
\begin{align}\label{Inverselaplce}
\mathcal{L}\left[x^{\beta-1}E_{\alpha,\beta}(-\lambda x^\alpha) \right](z) = \frac{z^{\alpha-\beta}}{z^\alpha + \lambda}.   
\end{align}

\subsubsection{Purely diffusive motion of heavy particle} 
To calculate $\langle p^2(t) \rangle$ for a Brownian particle undergoing purely diffusive motion ($\gamma = 0$) in 1D, I write (Eq. \eqref{Langevin_p}) as,
\begin{align}\label{Langevin_p_diff}
^{C}D^\alpha_{0+}p(t) = \xi(t),
\end{align}
  I use Eq. \eqref{sublaplaceofv} and Eq. \eqref{Langevin_p_diff} to solve it further as,

\begin{widetext}
\begin{align*}
     p^2(t) =&\;  \left(p(0) + tp^\prime (0) + \frac{1}{\Gamma(\alpha)}\int_0^t (t-s)^{\alpha-1}\xi(s)ds \right)^2\\
     =&\;p^2(0) + 2t p(0)p^\prime(0) + \frac{2p(0)}{\Gamma(\alpha)}\int_0^t (t-s)^{\alpha -1}\xi(s)ds\\
    & + t^2(p^\prime(0))^2 + \frac{2tp^\prime(0)}{\Gamma(\alpha)} \int_0^t (t-s)^{\alpha -1}\xi(s)ds+\left(  \frac{1}{\Gamma(\alpha)}\int_0^t (t-s)^{\alpha -1}\xi(s)ds \right)^2.
\end{align*}

Now calculating $\langle p^2(t) \rangle$ by using the properties of the white noise correlation as defined in Eq. \eqref{corr} and Taking initial conditions $p(0)=0$, and $p^\prime(0)=0$,

\begin{align}\label{momentum_analytic}
\langle p^2(t) \rangle\nonumber&= \frac{1}{\Gamma(\alpha)^2} \int_0^{t} \int_0^{t} (t-s_1)^{\alpha -1}  (t-s_2)^{\alpha -1} \langle \xi(s_1) \xi(s_2) \rangle ds_1 ds_2 \\
\nonumber& = \frac{1}{\Gamma(\alpha)^2} \int_0^{t} \int_0^{t} (t-s_1)^{\alpha -1}  (t-s_2)^{\alpha -1} 2\mathcal{D} \delta(s_1 -s_2) ds_1 ds_2 \\
\nonumber& = \frac{2\mathcal{D}}{\Gamma(\alpha)^2} \int_0^t (t-s)^{2\alpha -2}  ds \\
& = \frac{2\mathcal{D}}{\Gamma(\alpha)^2} \frac{t^{2\alpha-1}}{2\alpha-1}.
 \end{align}

In the similar way one can calculate the  $\langle x^2(t) \rangle$  with initial conditions, $x(0)=0$, and $x'(0)=0$ from Eq.~\eqref{sublaplaceofX}, as

\begin{align}\label{position_analytic}
   \langle x^2(t)  \rangle &\approx \frac{2\mathcal{D}}{\Gamma(\alpha+\beta)^2} \int_0^{t}  (t-s)^{2\alpha+2\beta -2}ds  =\frac{2\mathcal{D}}{\Gamma(\alpha+\beta)^2} \frac{t^{2\alpha+2\beta-1}}{2\alpha+2\beta-1}.
\end{align}
\end{widetext}

\begin{figure*}
		\centering
        \includegraphics[height=7.5cm,width=8.cm]{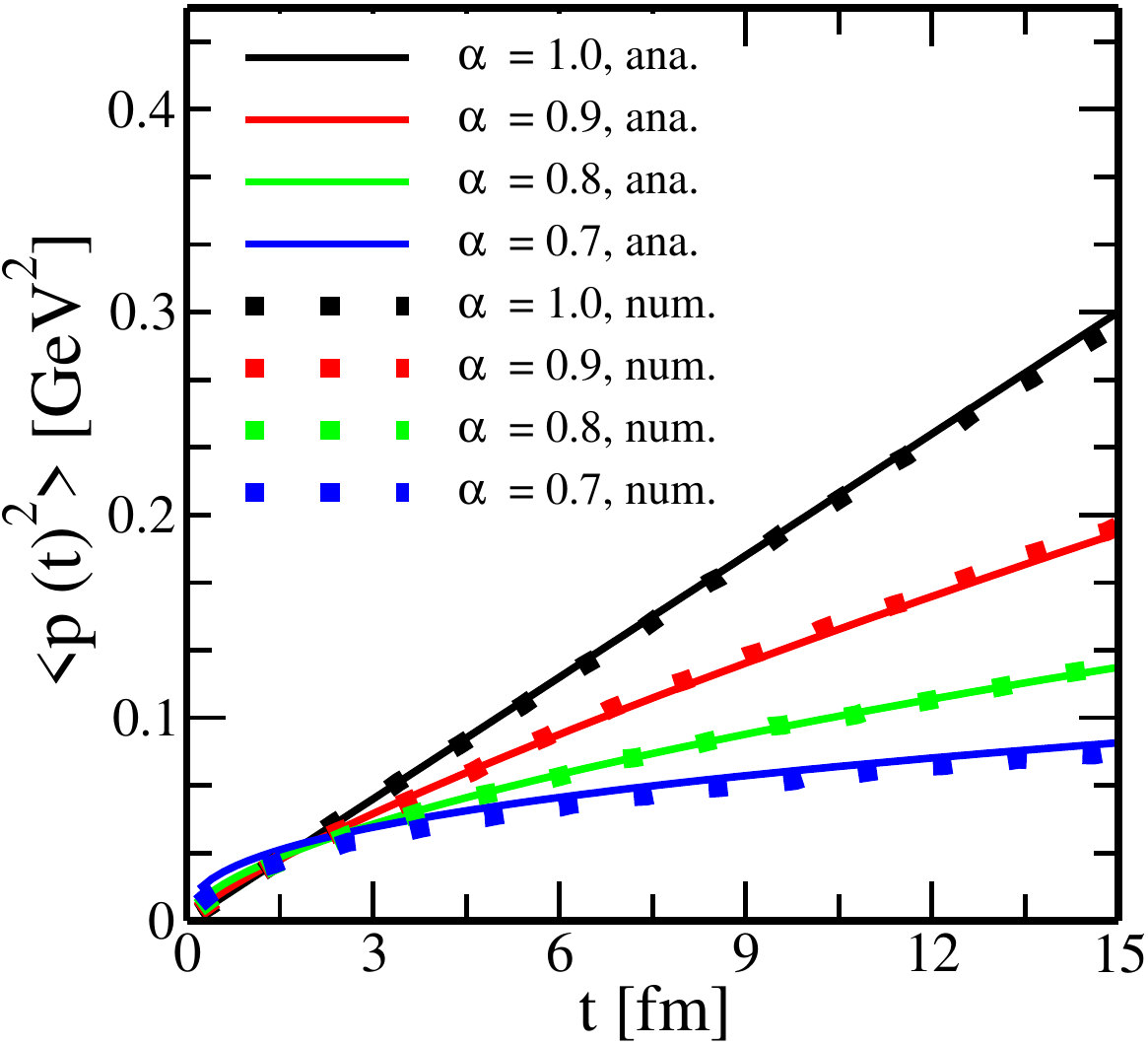}
        \hspace{10mm}
		\includegraphics[height=7.5cm,width=8.cm]{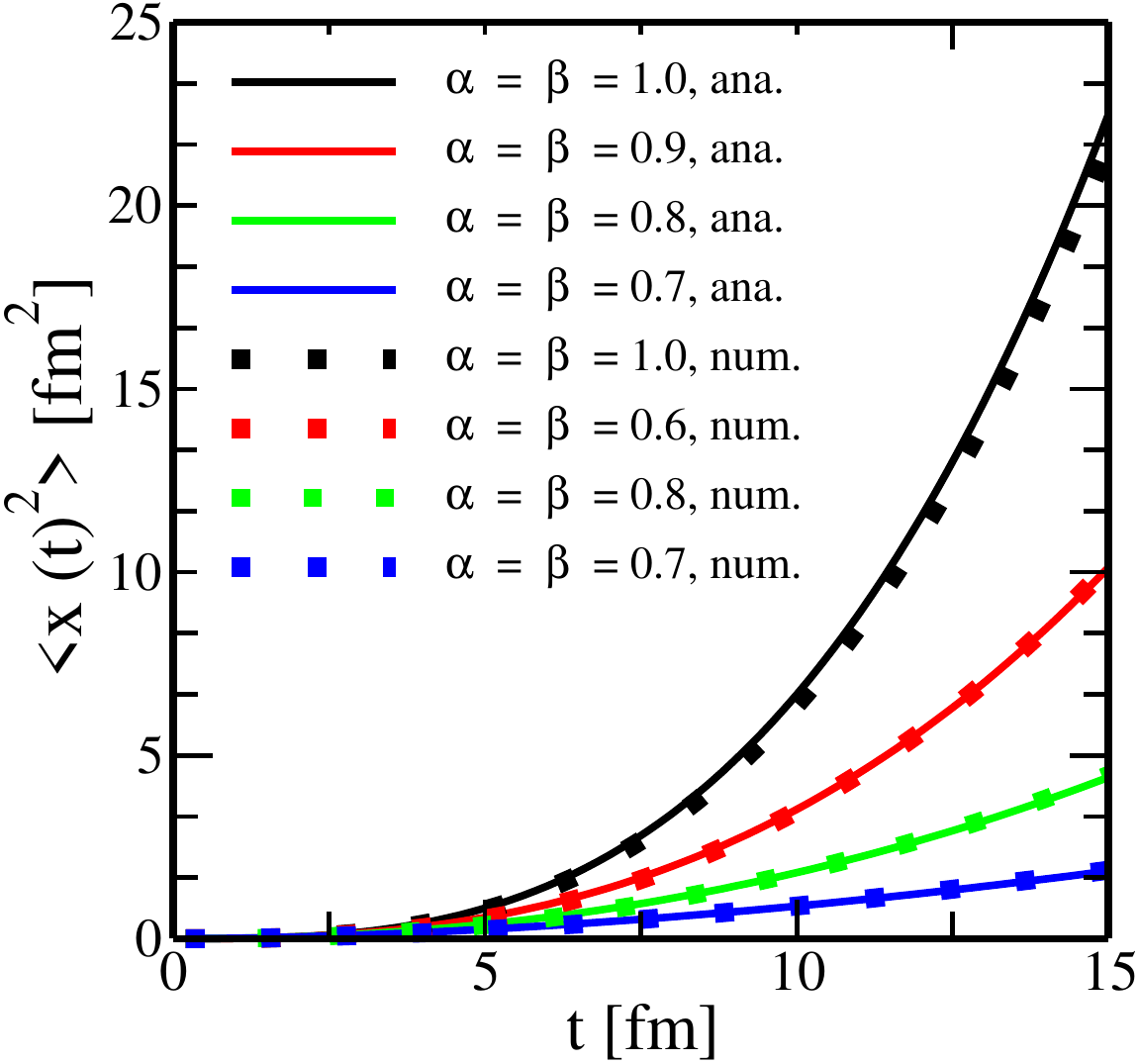}
		
		\caption{$\langle p^2(t) \rangle$ (left panel)  and $\langle x^2(t) \rangle$  (right panel) versus time for subdiffusion. These graphs consider a 1D purely diffusive motion, constant $\mathcal{D}$ of 0.1 GeV$^2$/fm and { $M$ = 1} with different values of $\alpha$ and $\beta$.}
		\label{x_1D}
	\end{figure*}
In fig.~\ref{x_1D}, I present the analytical solutions (represented by all solid lines) for $\langle p^2(t) \rangle$ and $\langle x^2(t)\rangle$, from in Eqs. \eqref{momentum_analytic}, \eqref{position_analytic}, respectively.
In Fig. \ref{x_1D} (left panel), I show the variation of $\langle p^2(t) \rangle$ over time in purely diffusive ($\gamma = 0$) motion, plotted for various values of $\alpha$, specifically 1.0 (black), 0.9 (red), 0.8 (green), and 0.7 (blue), at a constant $\mathcal{D}$ of 0.1 GeV$^2$/fm. For $\alpha$ $<$ 1, the time evolution of $\langle p^2(t) \rangle$ shows a transition from normal diffusion to subdiffusion. For $\alpha =1$ (corresponding to normal diffusion), $\langle p^2(t) \rangle$ increases linearly over time, reflecting a consistent rate of momentum spread. For $\alpha$ = 0.9, 0.8, and 0.7, the rate of increase in $\langle p^2(t) \rangle$ becomes progressively slower, indicating a subdiffusive behavior where the momentum spread is less efficient. This slower growth in $\langle p^2(t) \rangle$ with decreasing $\alpha$ highlights the reduced efficiency of momentum transfer, characteristic of subdiffusion for the Brownian particle. 

The right panel of  Fig. \ref{x_1D} presents $\langle x^2(t) \rangle$ over time for various values of $\alpha$, $\beta$, specifically 1.0 (black), 0.9 (red), 0.8 (green), and 0.7 (blue). Using similar simplification and the same constant value of the parameter. {In this case, I set $\gamma = 0$ in the FLE, where diffusion dominates the behavior of the Brownian particle. As a result, $\langle x(t)^2\rangle$ for the non-relativistic Brownian particle exhibits a time dependence with an exponent greater than one, deviating from the expected behavior of normal diffusion described by Eq. \eqref{MSD}. To address this discrepancy, I will include the drag term in the relativistic Langevin equation for the HQs in the next section.} For the lower values of $\alpha$ and $\beta$, 0.9, 0.8, and 0.7, the rate of increase in $\langle x^2(t) \rangle$ slows down, showing subdiffusive behavior where the spread of the particle's position becomes less efficient over time. The graph shows that for lower values of $\alpha$ and $\beta$, the growth in $\langle x^2(t) \rangle$ is markedly slower. In the following subsection, I solve the FLE using a numerical method. 
\subsection { Numerical method for the fractional Langevin equation}

The It\^o or Stratonovich stochastic calculus is typically employed for solutions in classical stochastic differential equations driven by Brownian motion. However, these methods are not suitable for the FLE, as it is not a semimartingale~\cite{rogers1997arbitrage}. Although the Monte Carlo method is a trustworthy approach for solving stochastic differential equations, it is not suitable in our article due to its reliance on independent sequences, while fractional noise sequences are dependent. Therefore, I have utilized L1 numerical schemes, as detailed in \cite{Jin2015AnAO,LIN20071533}. These methods are among the most effective for discretizing the Caputo fractional derivative and represent a key contribution to this paper. Numerical methods for solving the FLE are commonly categorized as indirect or direct. Time-fractional differential equations can typically be reformulated into integro-differential equations corresponding to indirect methods. In contrast, direct methods focus on approximating the time-fractional derivative itself. In this article, I aim to discretize the fractional derivative directly without transforming the associated differential equation into its integral form. Our numerical algorithm for solving the FLE is shown in Appendix \ref{AP}.

\subsubsection{The FLE for heavy quark: QGP}
I expanded the solution of the FLE of the non-relativistic Brownian particle  (Eqs.~\eqref{Langevin_x} \eqref{Langevin_p}) to describe the dynamics of the relativistic HQs in the QGP medium in the relativistic limit. The FLE for the momentum and the position evolution of the HQs is written as,

\begin{align}\label{Langevin_x_rel}
&^{C} D^\beta_{0+}x(t)=\frac{p(t)}{E(t)},
\\
&^{C} D^\alpha_{0+}p(t) = -\gamma p(t)+\xi(t),
\label{Langevin_p_rel}
\end{align}
where $p(t)$ denotes the momentum of the HQs and  $E = \sqrt{p^2+ M^2}$ denotes the energy. The $\gamma$ can be related to the diffusion coefficient through an FDT as \cite{PhysRevLett.84.31, Moore:2004tg,Mazumder:2013oaa}, 
\begin{align}\label{FDT_rel}
\gamma = \frac{\mathcal{D}}{ET},
\end{align}
given that both Eq.~\eqref{Langevin_x_rel} and Eq.~\eqref{Langevin_p_rel} are nonlinear differential, integral equations; because of this nonlinearity, the analytical solution through traditional methods such as Laplace transformations is not possible, as applied to their non-relativistic Eqs.~\eqref{Langevin_x} and \eqref{Langevin_p}. Consequently, solving the relativistic FLE is feasible exclusively via numerical simulations. I write the numerical code for the relativistic FLE (Eqs. \eqref{Langevin_x_rel}, \eqref{Langevin_p_rel}). The numerical form is described in the appendix \ref{AP}.
One can write using the discrete form by using Eq.~\eqref{AP_sub} (using L1 numerical scheme for the subdiffusion \ref{AP}), as follows,  
\begin{widetext}
\begin{align}\label{SubA_x}
\begin{cases}
  x(t_1)= x(t_0) + \left[\displaystyle\frac{p(t_0)}{E(t)}\right]k_1  \; &:\;n=1, \\
  x(t_n)= x(t_{n-1})+ \Bigg[\displaystyle\frac{p(t_{n-1})}{E(t)}  
 -\sum_{j=1}^{n-1} a_j \left(x(t_{n-j})-x(t_{n-j-1}))\right)\Bigg]k_1 \; &:\;n \geq 2, 
 \end{cases}
\end{align}
and,
\begin{align}\label{SubA_p}
\begin{cases}
 p(t_1)= p(t_0) + \left[-\gamma p(t_0)+\sqrt{2\mathcal{D}/\Delta{t}}~\eta(t)\right]k_1  \; &:\;n=1, \\
 p(t_n)= p(t_{n-1})+ \Bigg[-\gamma p(t_{n-1})+\sqrt{2\mathcal{D}/\Delta{t}}~\eta(t) 
 -\sum_{j=1}^{n-1} b_j\left(p(t_{n-j})- p(t_{n-j-1})\right) \Bigg]k_1 \; &:\;n \geq 2, 
 \end{cases}
\end{align}
\end{widetext}

where
$k_1 = \Delta t^{\alpha}\Gamma(2-\alpha)$,   $t_{n}= t_{n-1} +\Delta t$,  $t_{0}$ is the initial time, and $\Delta{t}$ is the time step.

\subsection{ A check for numerical scheme}

 I calculate numerically $\langle x^2(t) \rangle$ from Eq.~\eqref{SubA_x} and $\langle p^2(t) \rangle$ from Eq.~\eqref{SubA_p} (replacing the energy  by the mass of the Brownian particle). Thereafter, the numerical results are compared with the analytical solutions calculated from Eq.~\eqref{position_analytic} and Eq.~\eqref{momentum_analytic}, respectively.  This step is important to confirm the accuracy of the L1 numerical method. The FLE has been solved in the non-relativistic limit, as detailed in Sec.~\ref{NONRELATIVISTIC LIMIT}. For illustration, I have assumed a constant diffusion coefficient $\mathcal{D}$ of 0.1 GeV$^2$/fm, a particle mass $M$ of 1 GeV, and $\gamma = 0$ (representing a pure diffusion case) in the FLE.

In  Fig. \ref{x_1D} (left panel), $\langle p^2(t) \rangle$ is estimated for various different $\alpha$, specifically 1.0 (black), 0.9 (red), 0.8 (green), and 0.7 (blue). Notably, these numerical results (represented by dashed lines) match with the analytic solutions. To further validate our numerical method and ensure the precision and significance of our numerical approach. I numerically calculated $\langle x^2(t) \rangle$. The results are displayed in Fig. \ref{x_1D}  (right panel). $\langle x^2(t) \rangle$ is calculated for $\alpha$ and $\beta$ values of 1.0 (black), 0.9 (red),0.8 (green), and 0.7 (blue). Lower values of $\alpha$ and $\beta$ are not included in the provided plots because their magnitudes are significantly smaller and not noticeable on the same scale as Fig.  \ref{x_1D}. For the same reason, I have excluded results for lower $\alpha$ values in the following subsection. One can see that the numerical simulation agrees with the analytical result, signifying that our numerical scheme performs correctly. In the following section, I calculate the effect of subdiffusion on the HQ dynamics in the hot QCD matter.

\section{Results}
\label{Res}
In this section, I calculate the $\langle p^2(t) \rangle$, $\langle x^2(t) \rangle$, kinetic energy, and Normalized momentum correlation function of the HQs. The mass of the charm quark, $M_c$ = 1.3 GeV and for the bottom quark, $M_b$ = 4.5 GeV, are considered in this article. First, I assume a simplified initialization and constant $\mathcal{D}$ for illustrative purposes. Further, I calculate the $R_{AA}$, and $dN/dp_T$ to show the impact of subdiffusion on the HQs dynamics, where the transport coefficient is calculated from pQCD for elastic processes from the well-established literature ~\cite{Svetitsky:1987gq} for two different value of $T$. {I am using the relativistic FDT theorem to calculate $\gamma$, as defined in Eq. \eqref{FDT_rel}. This theorem is used in the calculations for Fig. 2 to Fig. 7}. The numerical results in this section are calculated using Eqs. \eqref{SubA_x}, \eqref{SubA_p}.

\subsection{The time evolution of $\langle p^2(t)\rangle$ and $\langle x^2(t)\rangle$ of the HQs in the QGP medium.}

In this article, I defined $\langle p^2(t) \rangle$ and $\langle x^2(t) \rangle$  as follows, 
\begin{align} \langle p^2(t) \rangle = 
  \langle p_X^2(t) + p_Y^2(t) \rangle, 
\end{align} 
and,
\begin{align} \langle x^2(t) \rangle = X^2(t) + Y^2(t), \end{align}

where $p_X(t)$ is the momentum in $X$ direction and $p_Y(t)$ is the momentum in $Y$ direction. I have calculated $\langle p^2(t) \rangle$ and $\langle x^2(t) \rangle$ of the HQs in the hot QCD matter for various values of $\alpha$ and $\beta$, namely 1.0 (black), 0.9 (red), 0.8 (green), and 0.7 (blue)  at $T = 250 $ MeV, and $\mathcal{D} = 0.1$ GeV$^2$/fm. The corresponding results are shown in Fig.~\ref{p_2D} and Fig.~\ref{x_2D} for the $\langle p^2(t) \rangle$ and $\langle x^2(t) \rangle$ respectively. In this case, the initial momentum is set to be $p_x(t_0) = p_y(t_0) = 0$ and initial position $X(t_0) = Y(t_0) = 0$. 
Fig. \ref{p_2D} shows that when $\alpha$ $<$ 1, the process behavior indicates subdiffusion for the charm quark ($M_c = 1.3 \text{GeV}$) within the hot QCD medium. The subdiffusive behavior becomes more pronounced at lower $\alpha$ values, demonstrating that the $\langle p^2(t) \rangle$ of the HQs grows more slowly due to the complex interactions within the hot QCD medium. This transition underscores the influence of $\alpha$ on the diffusion process, where lower $\alpha$ values reflect a deviation from normal diffusion, highlighting the subdiffusive dynamics of the HQs in the QGP medium. The plot for the bottom quark is not shown since I observed a similar behavior as charm quark.  As $\alpha$ approaches 1, the subdiffusion process converges to normal diffusion. Also, in the later stages, $\langle p^2(t) \rangle$ tends to approach $3MT$ \cite{Moore:2004tg}.

\begin{figure}
		\centering
        \includegraphics[height=7.5cm,width=8.cm]{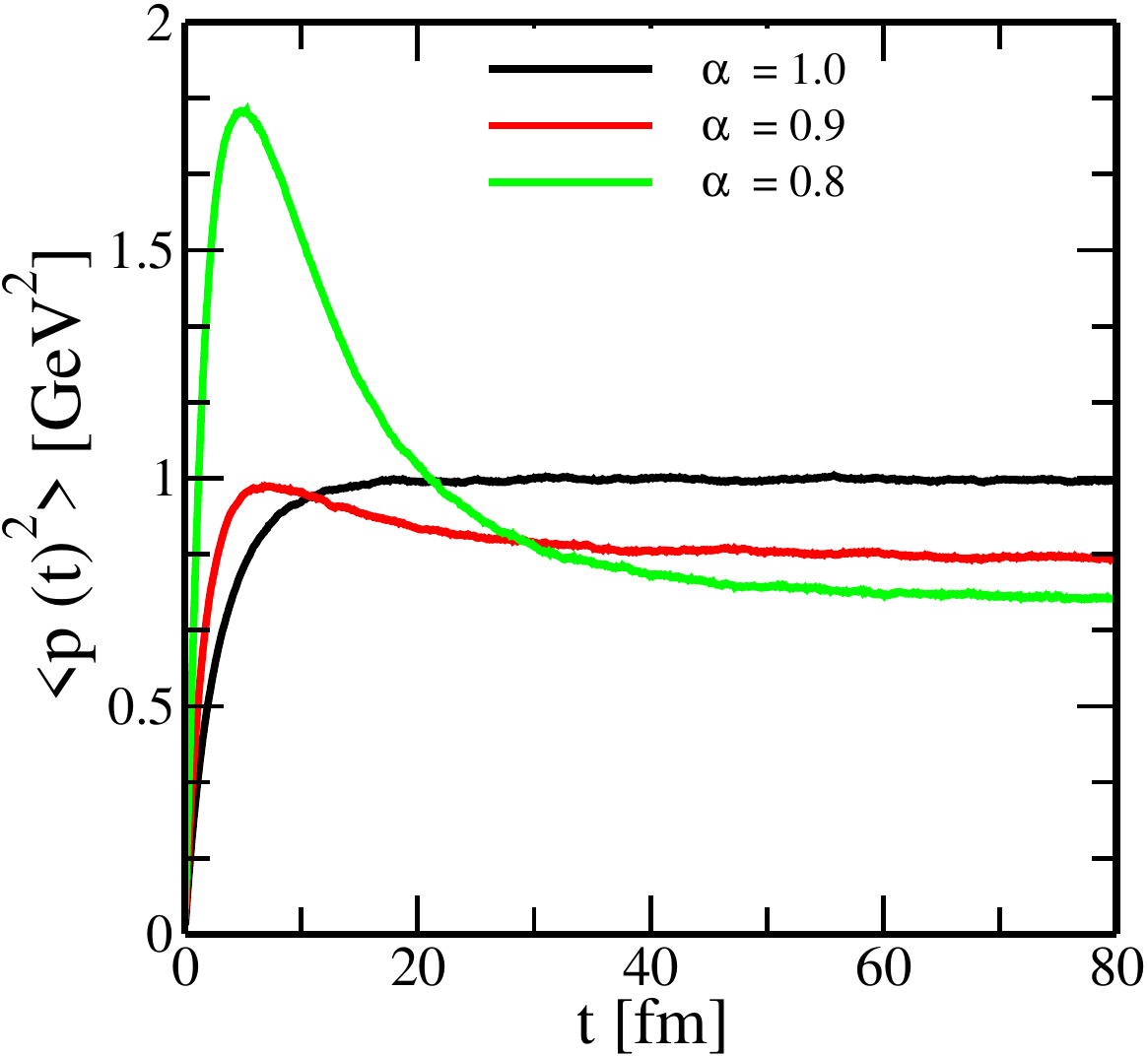}
\caption{$\langle p^2(t) \rangle$ versus time of the HQs subdiffusion with the various values of the $\alpha$ at a constant $\mathcal{D}$ of 0.1 GeV$^2$/fm and at $T = 250 $ MeV.}
		\label{p_2D}
	\end{figure}
 
 \begin{figure}
		\centering  
        \includegraphics[height=7.5cm,width=8.cm]{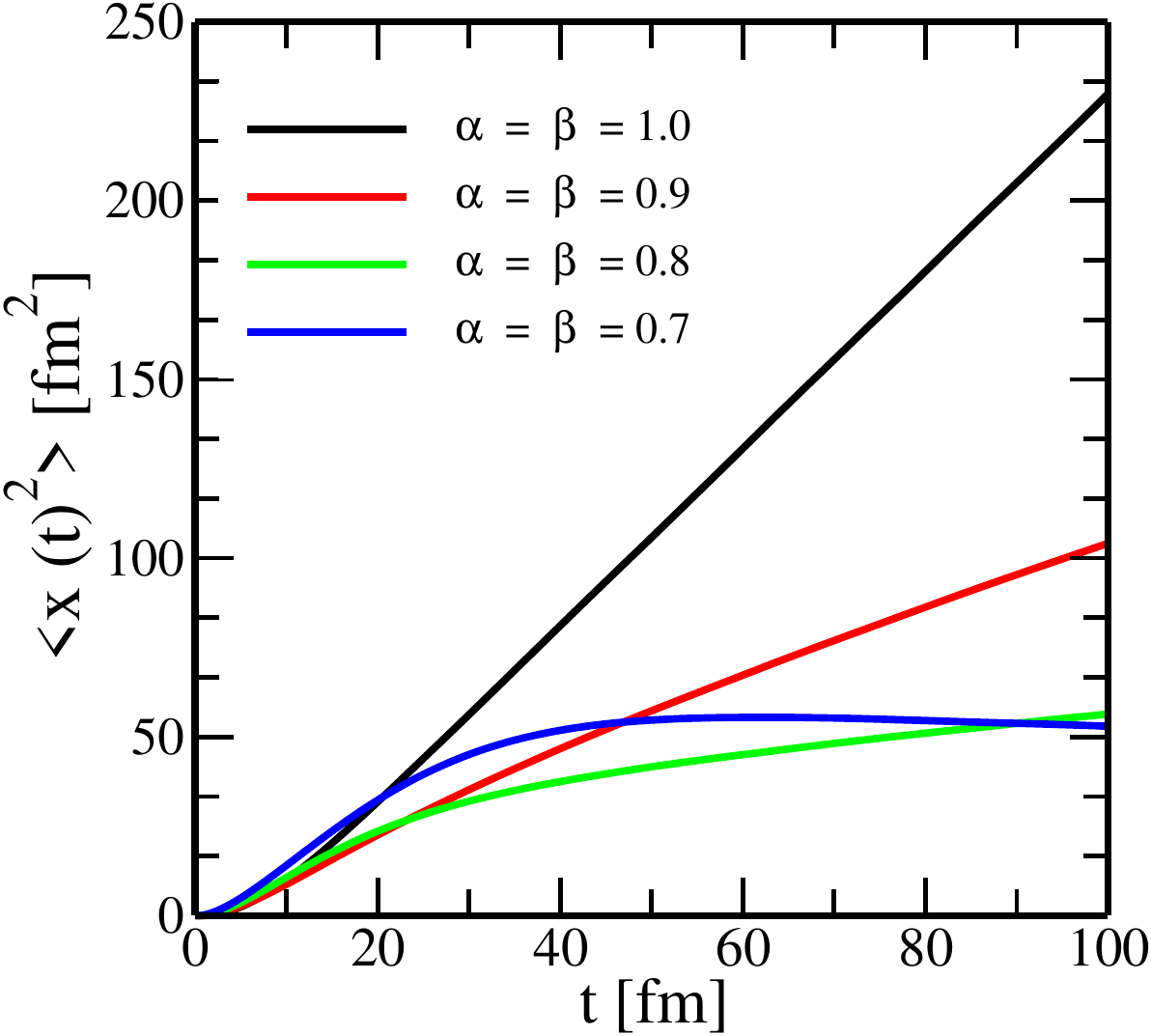} 
        \hspace{10mm}
\includegraphics[height=7.5cm,width=8.cm]{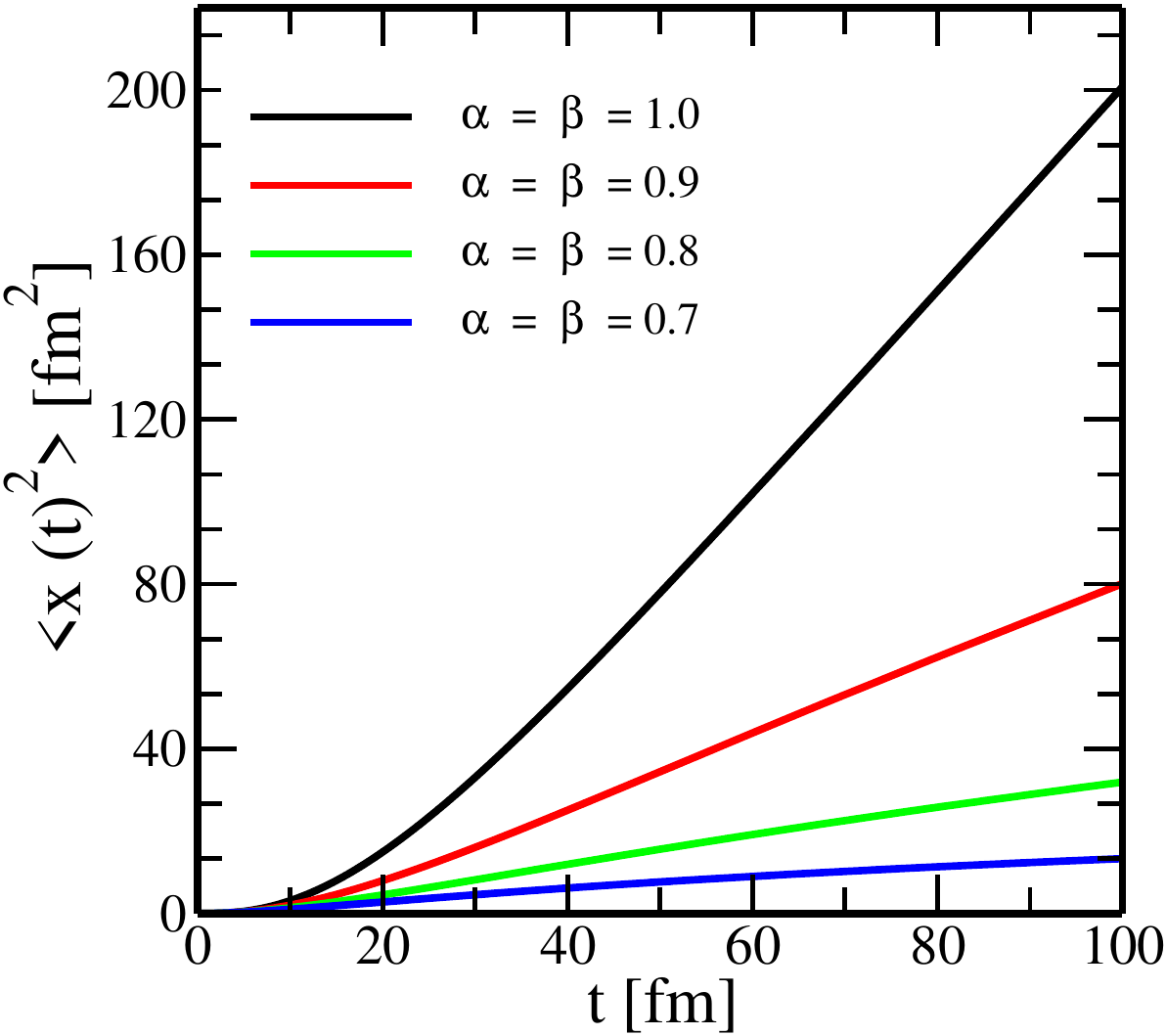} 
  \caption{$\langle x^2(t) \rangle$ (upper panel {for charm quark}) and $\langle x^2(t) \rangle$  (lower panel {for bottom quark}) versus time for subdiffusion, with the various values of the $\alpha$, $\beta$  at $\mathcal{D}$ = 0.1 GeV$^2$/fm and $T = 250 $ MeV.}
		\label{x_2D}
	\end{figure}

In Fig.~\ref{x_2D} (upper panel), I have calculated the evolution of $\langle x^2(t) \rangle$  for charm quark over time and in Fig.~\ref{x_2D} (lower panel) for bottom quark ($M_b$ = 4.5 GeV) for various values of $\alpha = \beta$ = 1.0 (black), 0.9 (red), 0.8 (green), and 0.7 (blue) maintaining other parameters consistent with those illustrated in Fig.~\ref{p_2D}. For both quarks, as $\alpha$ and $\beta$ decrease, the rate of increase in $\langle x^2(t) \rangle$ diminishes, indicating a transition from normal diffusion to subdiffusion. For $\alpha = \beta = 1.0$, $\langle x^2(t) \rangle$ increases linearly with time, characteristic of normal diffusion. In this limit, at a later time evolution, $\langle x^2(t) \rangle$ shows a proportional relationship with $t$ {(as discussed in Eq. \eqref{MSD})}, as illustrated in \cite{Moore:2004tg, Svetitsky:1987gq}. For $\alpha$ and $\beta$ $<1$, say 0.9, 0.8, and 0.7, the curves for both quarks show slower growth in $\langle x^2(t) \rangle$, reflecting subdiffusive behavior. However, the bottom quark's $\langle x^2(t) \rangle$ values are consistently lower than those of the charm quark at equivalent times and parameter values.
{This difference in the magnitude of $\langle x^2(t) \rangle$ suggests that the bottom quark experiences more pronounced subdiffusive behavior compared to the charm quark. The analysis indicates that both the charm and bottom quarks exhibit subdiffusion but with different patterns in $\langle x^2(t) \rangle$.} The bottom quark's motion is more hindered, emphasizing the impact of the mass and medium interactions on the diffusion process. This puzzling relation of $\langle p^2(t) \rangle$ and $\langle x^2(t) \rangle$ of the HQs due to the subdiffusion, this dynamics is not explained before in the context of the HQ dynamics in a hot QCD medium.

\subsection{ Normalized momentum correlation function}

\begin{figure}
		\centering
        \includegraphics[height=7.5cm,width=8.cm]{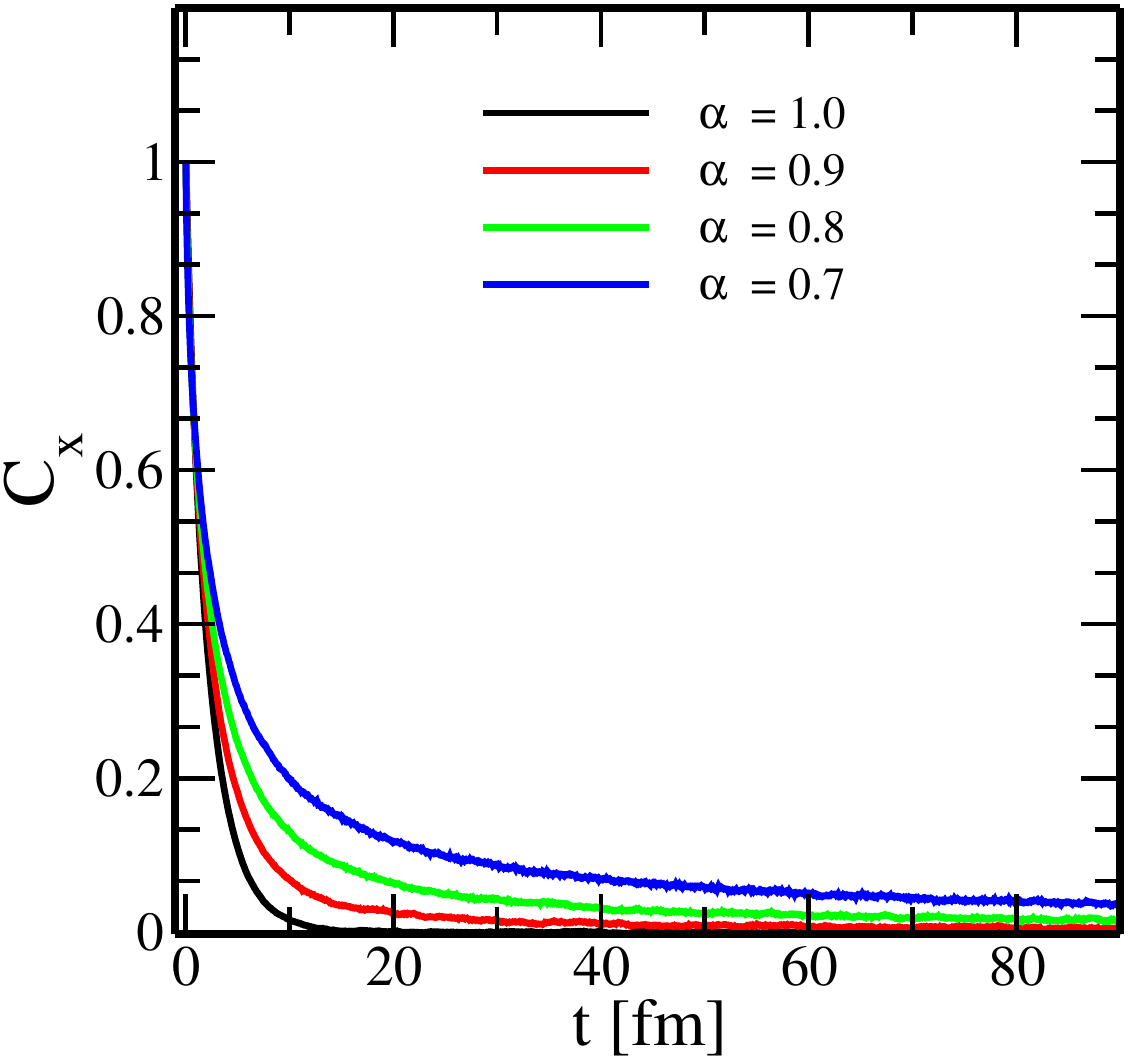}
        \caption{$C_x$ versus time of the HQs subdiffusion, considering the various values of the $\alpha$  and constant $\mathcal{D} = 0.2$ GeV$^2$/fm.}
		\label{correlation_p}
	\end{figure}

Normalized momentum correlation function ($C_x$) can be defined as follows,
\begin{align}
    C_x = \frac{\langle p_X(t)p_X(0)\rangle}{(p_X(0))^2},
\end{align}
where $C_x$ calculates the correlation between the momentum of the HQs in the $x$ direction at a given time and its momentum at a later time. The results for $C_y$ and $C_z$ in $Y$ and $Z$ direction respectively are similar. In the non-relativistic limit, $C_x$ is proportional to the auto-correlation function of velocity \cite{WOS:000332486500022,WOS:000291660200006,SANDEV20113627}. This function allows us to study how particle interactions with the surrounding medium diminish the initial momentum correlations.
In our study, the initial momentum of the HQs is set to $p_X(0)$ = 0.5 GeV, and $p_X(t)$ representing the momentum at any given time $t$.
In Fig. \ref{correlation_p}, the $C_x$ for charm quark subdiffusion is shown for various values of $\alpha$ = 1.0 (black), 0.9 (red), 0.8 (green), and 0.7 (blue) and at constant $\mathcal{D} = 0.2$ GeV$^2$/fm. The $C_x$ exhibits a monotonously decaying behavior. A key observation from Fig. \ref{correlation_p} is that the $C_x$ decays more slowly for lower values of $\alpha$. This indicates that for smaller $\alpha$, the momentum correlations persist for longer periods, characteristic of subdiffusive behavior, where correlations do not decay as quickly as in normal diffusion (when $\alpha = 1.0$). I have observed the same pattern of the $C_x$ for the bottom quark.

\subsection{Kinetic energy of the heavy quarks}
To emphasize the effect of the subdiffusion on the HQ dynamics, I have calculated the average kinetic energy ($KE$) by using the expression
\begin{align}
    KE = \langle\sqrt{p^2+M_c^2} -    M_c \rangle,
\end{align}

using the simple initialization corresponding to $p_T = 0.5$  GeV and $p_z = 0$,  corresponds to the midrapidity region of realistic collisions at the value for  $T = 0.25$ GeV, $M_c = 1.3$ GeV and $\mathcal{D} = 0.1$ GeV$^2$/fm corresponds with the diffusion coefficient calculated within the pQCD. 

  Fig.~\ref{kinetic} shows the time evolution of the $KE$ for the  HQs at different values of $\alpha$ such as 1.0, 0.9, 0.8, and 0.7. For $\alpha = 1$, the $KE$ rapidly increases and then reaches a plateau, which means that the charm reaches the thermal equilibrium with its surrounding matter and indicates that the charm quarks are thermalizing faster. This rapid thermalization is characteristic of normal diffusion, where the interactions between the charm quarks and the medium lead to a quick equilibration of energy.
For $\alpha$ $<$ 1, say 0.9, 0.8, and 0.7, the curve shows a slower increase in kinetic energy and a more gradual approach to the plateau. This slower rise indicates subdiffusive behavior, where the charm quark experience more complex interactions with the medium, resulting in a prolonged thermalization process. The plots for the bottom quarks are not shown since they also follow a similar pattern.
The slower thermalization for lower $\alpha$ values can be attributed to the subdiffusive nature of the medium, which causes the quarks to retain their initial conditions for a longer time. 

\begin{figure}
		\centering
        \includegraphics[height=7.5cm,width=8.cm]{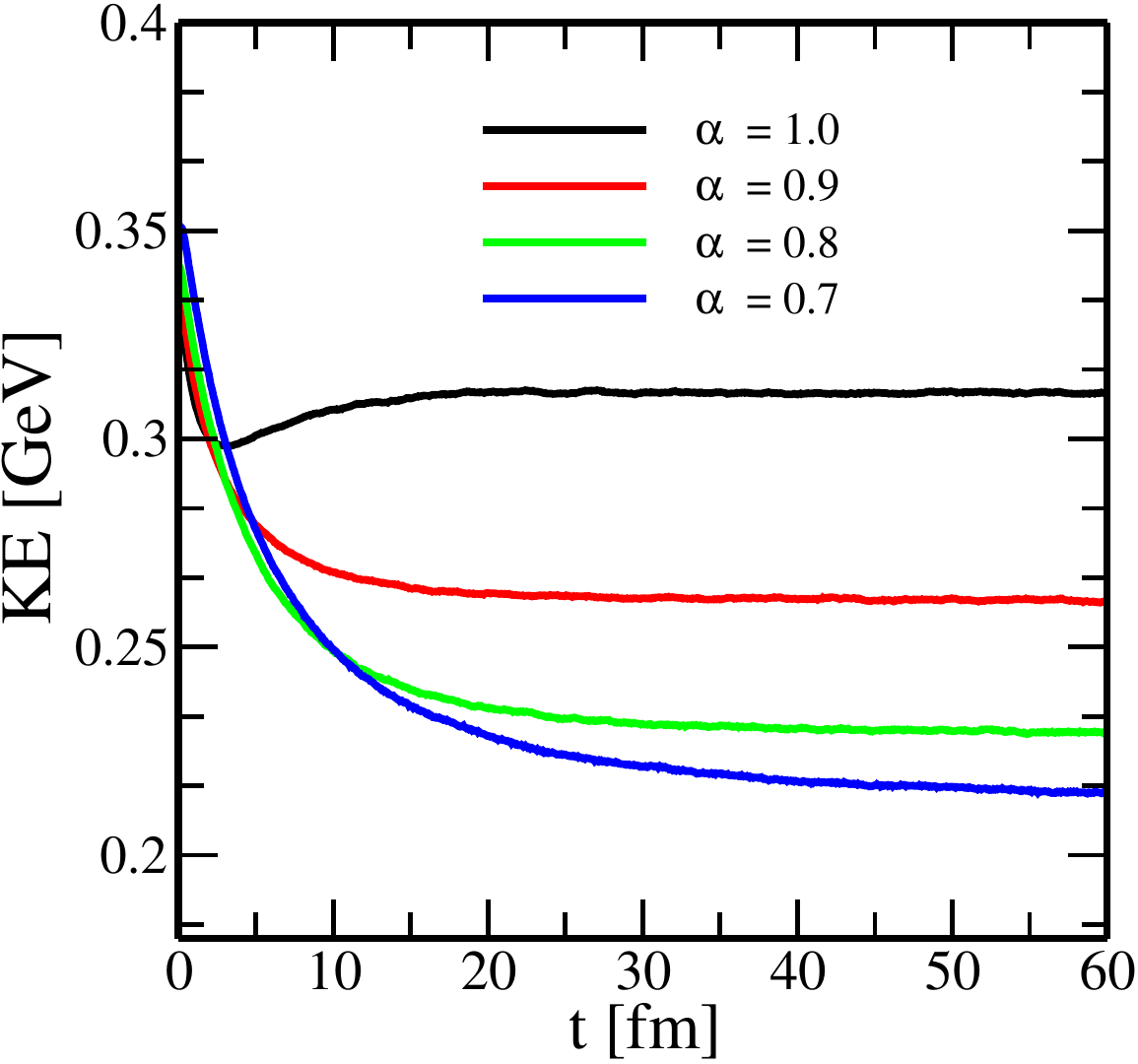}
        \caption{$KE$ versus time of the HQs subdiffusion, considering the various values of the $\alpha$  at fixed $\mathcal{D} = 0.1$ GeV$^2$/fm and $T = 250 $ MeV.}
		\label{kinetic}
	\end{figure}

\subsection{Momentum spread of HQs}

I calculate the evolution of $dN/dp_T$ of the bottom quark at static temperatures of 250 MeV (top panel) and 500 MeV (bottom panel) in Fig. \ref{momentum_distribution}.  The evolution of $dN/dp_T$ is shown for various values of $\alpha$ at a final evolution time of $\tau_f$ = 6 fm/c. Initially, all bottom quarks are concentrated within a very narrow $p_T$ bin, creating a delta-like distribution at $p (t_0)$ = 10 GeV (represented by the red line). As the bottom quarks interact with the QGP medium, the $dN/dp_T$ distribution spreads out.
For subdiffusion, $\alpha = 0.5$ (green line), $\alpha = 0.4$ (blue line), and $\alpha = 0.3$ (magenta line) are considered, the distribution $dN/dp_T$ shows a reduced spread compared to $\alpha = 1$, and the average momentum shifts towards higher $p_T$ values. Specifically, for $\alpha = 0.3$, the spread is minimal, and the average momentum shifts towards the highest $p_T$ values, indicating a slower thermalization process due to the subdiffusion. I have taken the combined effects of diffusion and drag coefficients of the bottom quarks. In contrast, for $\alpha = 1$, which corresponds to normal diffusion, the distribution spreads more widely, as explained in Refs. \cite{PhysRevC.84.064902, Das:2013kea}.
Despite the differences in the spread and shift of the distributions for different $\alpha$ values, the total area under each curve remains constant, reflecting the conservation of the number of the bottom quarks. The effect of subdiffusion is consistent at both temperatures. A similar pattern is observed for the charm quarks as well, so that is avoided to show in the article. To show the maximum impact of the subdiffusion, other higher values of $\alpha$ like 0.9, 0.8, 0.7 are not considered in the calculation of $dN/dp_T$ and for the $R_{AA}(p_T)$ in the next subsection.

\begin{figure}
		\centering
        \includegraphics[scale = .3]{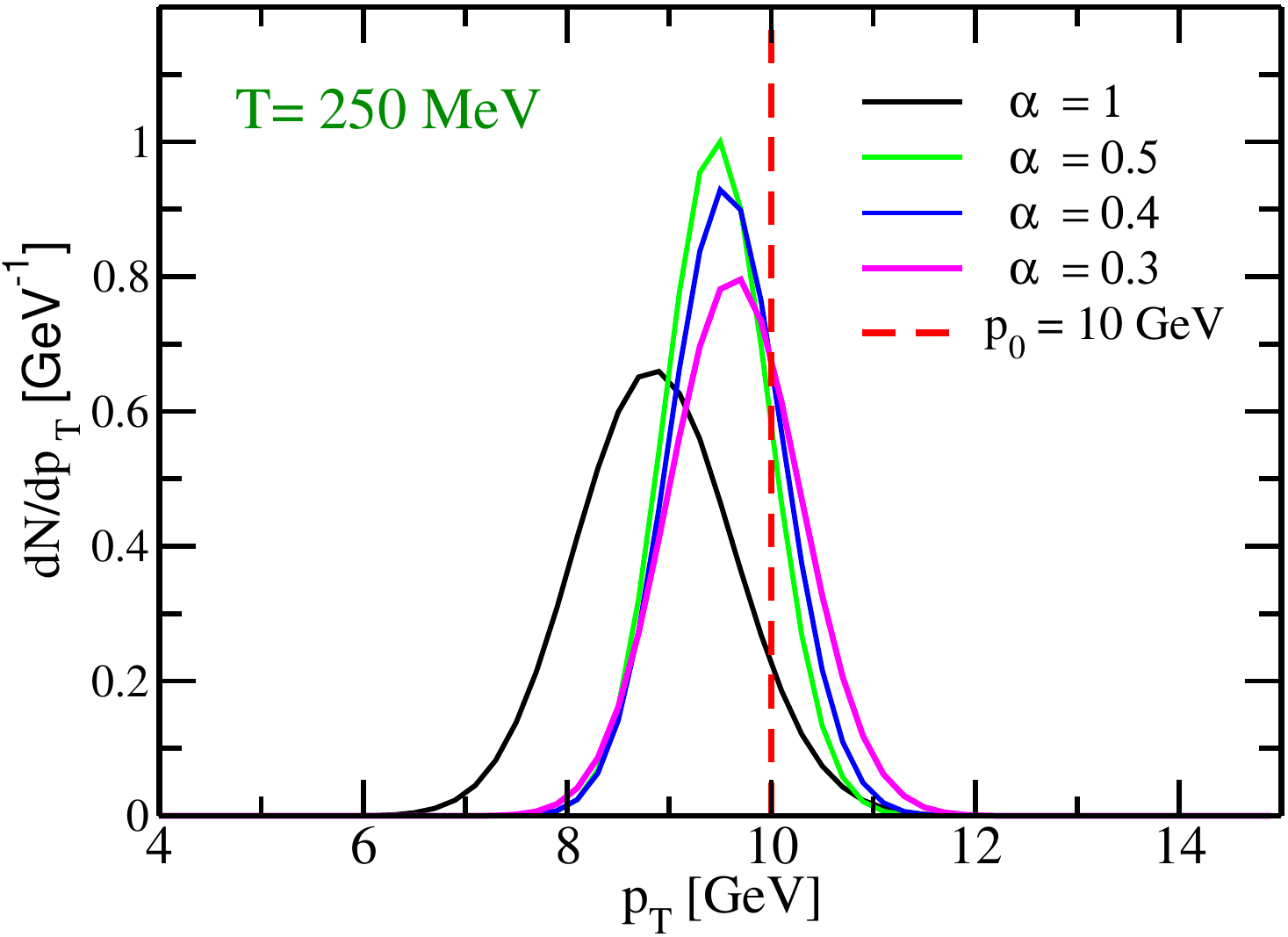}
         \hspace{10mm}
		\includegraphics[scale = .3]{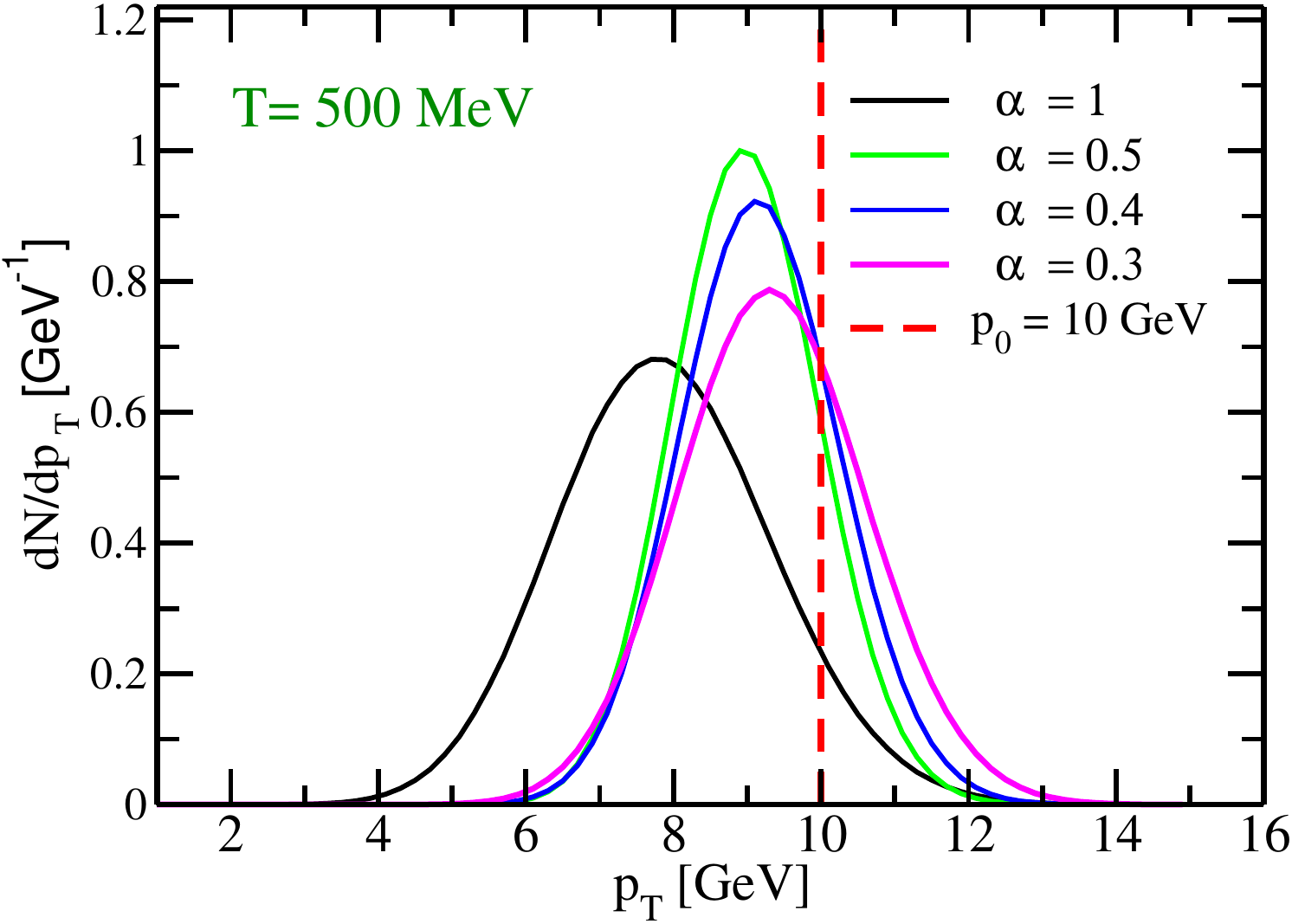}
        
		\caption{ The evolution of bottom quark momentum distribution as a function of $p_T$ at evolution time, $\tau_f$ = 6 fm/c, considering different $\alpha$ and at two different temperatures, $T$ = 250 MeV (above panel) and $T$ = 500 MeV (below panel). Assuming an initial delta distribution centred at $p(t_0)$ = 10 GeV.}
\label{momentum_distribution}
	\end{figure}

\subsection{Nuclear modification factor}

\begin{figure*}
		\centering
        \includegraphics[scale = .32]{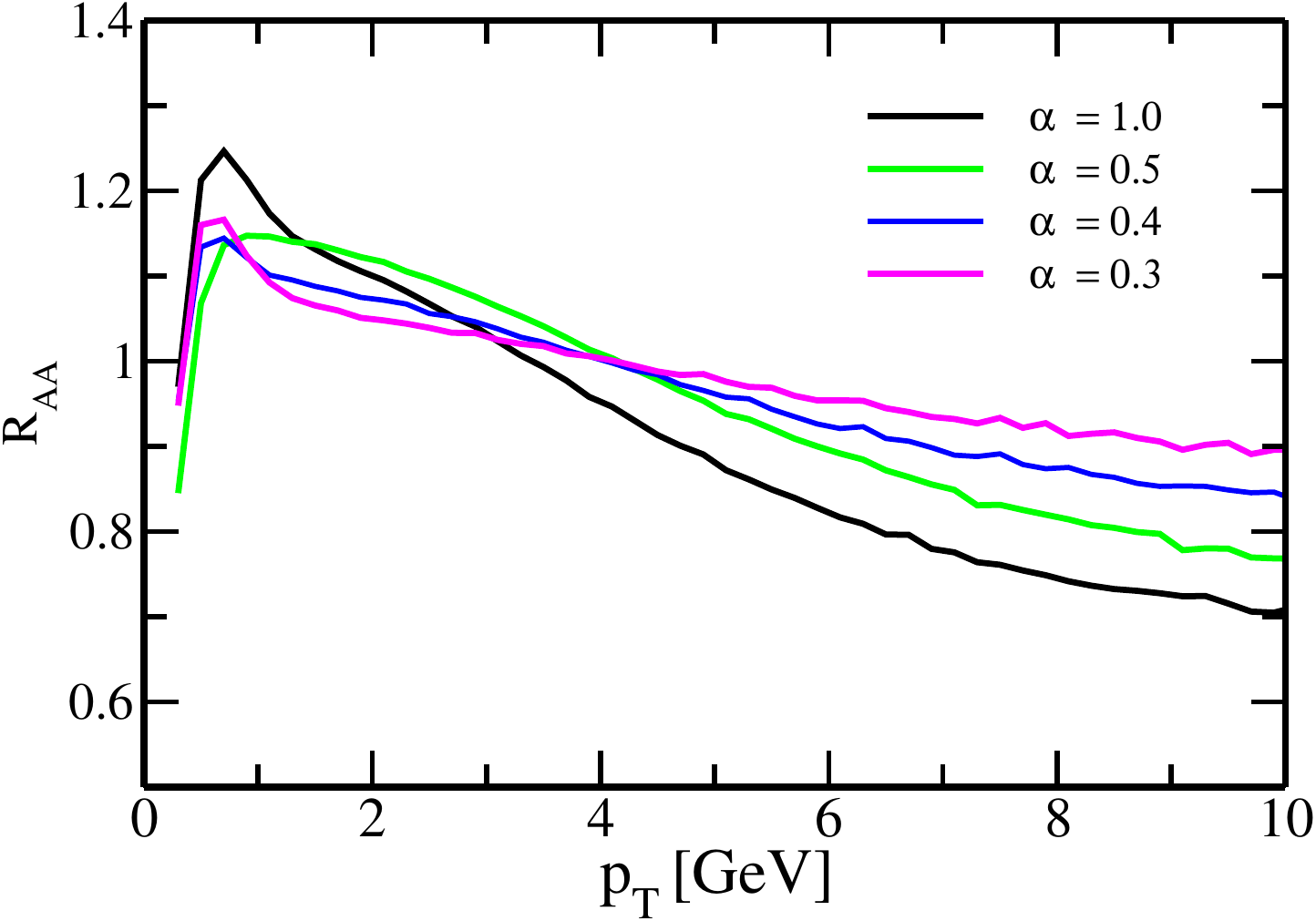}
         \hspace{10mm}
		\includegraphics[scale = .32]{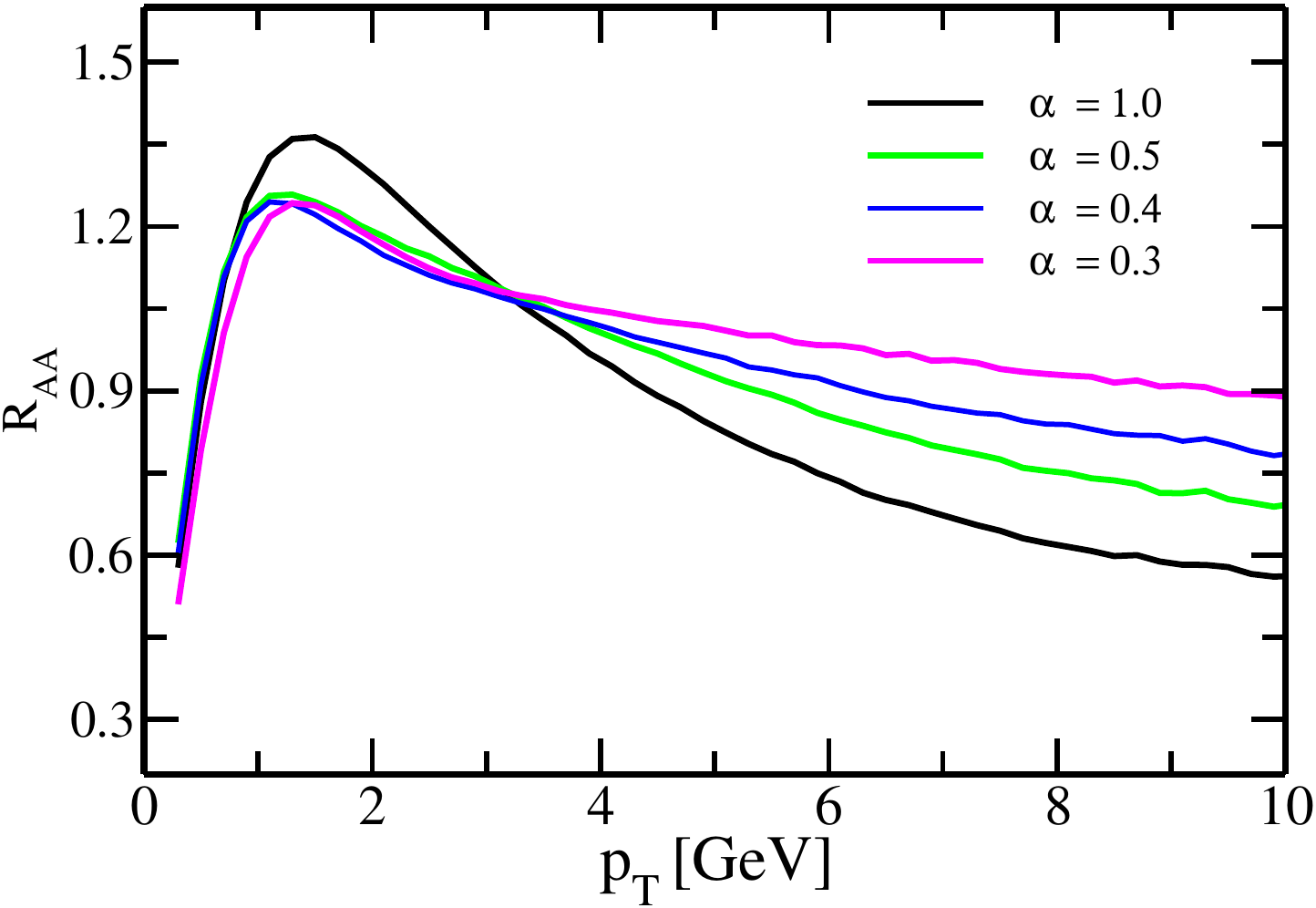}
        
		\caption{The $R_{AA}$ as a function of $p_T$ for the bottom quark, with $t$ = 6 fm/c, and at two different temperatures ($T$ = 250 MeV (left panel) and $T$ = {500} MeV (right panel) with various different values of  $\alpha$.}

		\label{raa}
	\end{figure*} 
To study the effect of the subdiffusion on the experimental observable, I have computed the  $R_{AA}(p_T)$ for the bottom quark, which is defined as follows \cite{Moore:2004tg},

\begin{align}
 R_{AA}(p_T)=\frac{f_{\tau_f} (p_T )}{f_{\tau_0} (p_T)},   \end{align}
the momentum spectrum,  $f_{\tau_f}(p)$, of the bottom quark estimated for the time evolution, $\tau_f$ = $6$ fm/c in our numerical results. $f_{\tau_i}(p_T)$ is the initial momentum distribution of the bottom quark, which  is taken according to the fixed order + next-to-leading log  calculations, which is capable of reproducing the spectra of B-mesons produced in proton-proton collisions through fragmentation~\cite{cacciari2005qcd, Cacciari:2012ny},  defined as,
 \begin{equation}
     \frac{dN}{d^2p{_T}} = \frac{x_0}{(1+x_3{p_T}^{x_1})^{x_2}},
\end{equation}
where the parameters are taken as follow;  $x_0= 0.46799$, $x_1= 1.83805$, $x_2= 3.07569$ and $x_3= 0.030156$. 
$R_{AA} (p_T)$ $\ne$ 1 implies that
HQs undergo interactions with the
medium. These interactions lead to modifications in the spectrum of charm quarks.

In Fig.~\ref{raa}, the  $R_{AA}$ is shown as a function of $p_T$ for various values of $\alpha$.
The calculations are conducted at $T$= 250 MeV (left panel) and $T$= 500 MeV (right panel). As $p_T$ increases, the magnitude of the $R_{AA}$ is decreasing (indicating more significant suppression at higher momenta.), which can be noticed for both $T$  in Fig.~\ref{raa}. For the subdiffusion, where the value of the $\alpha$ are considered as = 0.5 (green line), $\alpha = 0.4$ (blue line), and $\alpha = 0.3$ (magenta line). For $\alpha \rightarrow 1$ (depicted by the black line), the behavior of the $R_{AA}$  belong to normal diffusion, consistent with results available in the literature for the identical input parameters \cite{PhysRevC.93.014901, Das:2013kea,Das:2015ana}. I noticed that as the value of $\alpha$ is decreasing, then the formation of the $R_{AA}$ is slowdown. This effect is seen for both $T$.  The study also mentions that the effect of subdiffusion on charm quarks is similar to the results shown in the plot. However, the charm quark $R_{AA}$ plots are not displayed to maintain clarity and avoid overcrowding the figure.

\section{Conclusion and outlooks}

In this study, I studied the subdiffusion process of HQs in the QGP by using the FLE incorporating the Caputo fractional derivative. This article addresses the problem of the HQs in relativistic HICs, where HQs diffuse and lose energy within the QGP medium.
I solved the analytical and numerical FLE of nonrelativistic Brownian particles for the 1D motion for the $\gamma = 0$ using the Laplace transformation technique. Then, it showed that the numerical and analytic solutions for the $\langle x^2(t) \rangle$ and $\langle p^2(t) \rangle$ match exactly. The relativistic FLE cannot be solved analytically because of the nonlinearity. So, I solved the relativistic FLE numerically for the HQs dynamics in the hot QCD matter.
Our analysis revealed that the $\langle x^2(t) \rangle$ of the HQs follows a power-law time dependency, as illustrated in Fig. \ref{x_2D}. I found that as $\alpha$ and $\beta$ approaches 1, the system transitions to normal diffusion. Our numerical analysis highlighted the distinct characteristics of subdiffusion compared to normal diffusion. In the long-time limit, the HQs approach thermal equilibrium with the constituent particle of the QGP medium.

Key quantities such as $\langle p^2(t) \rangle$, $\langle x^2(t) \rangle$,  $C_x$, KE, $dN/dp_T$ and finally $R_{AA}$ were calculated to describe the dynamics of the HQs undergoing subdiffusion. Our results showed that subdiffusion, characterized by lower $\alpha$ values, leads to slower spreading of the momentum distribution and longer-lasting correlations in the momentum. For example, in Fig. \ref{p_2D}, the mean squared momentum $\langle p^2(t) \rangle$ shows a slower increase for lower $\alpha$ values, indicating subdiffusive behavior. Similarly, Fig. \ref{x_2D} shows that the $\langle x^2(t) \rangle$ grows more slowly as $\alpha$ and $\beta$ decreases.
The kinetic energy evolution, as shown in Fig. \ref{kinetic}, indicated that subdiffusion results in slower thermalization compared to normal diffusion ($\alpha$ = 1). Additionally, the momentum distribution $dN/dp_T$ at different static temperatures (Fig. \ref{momentum_distribution}) showed a reduced spread for lower $\alpha$ values, with the average momentum shifting towards higher $p_T$ values, reflecting the impact of subdiffusion. 
{ The proposed FLE framework offers an effective method for describing the HQs dynamics without focusing on the microscopic details of interactions such as elastic scattering and in-medium gluon emission, which are already well understood. Our approach may improve the description of radiative energy loss of the HQs in the $R_{AA}$. However, I acknowledge that this method may not address phenomena like the HQs thermalization due to the loss of the Einstein relation. I see the potential for future work to incorporate explicit radiative energy loss terms.} 

{Our study is primarily theoretical, and while it presents an unexplored method for modelling the HQs dynamics within the framework of the FLE, it remains to be seen whether this new approach is an improvement over current models. This conclusion will require further validation with experimental data. Such comparisons are crucial for establishing the practical advantages and reliability of the FLE framework.}

Our current study may give rise to thinking about the combined effect of time-correlated noise and subdiffusion on the HQ motion in the hot QCD matter.  This study represents a significant step towards understanding the complex dynamics of heavy quarks in QGP and highlights the potential of subdiffusion as a more comprehensive framework for describing their dynamics. This work applies the basis for a more realistic diffusion model of HQs, which should incorporate a proper initial geometry and an expanding medium. In the future, it will be essential to explore whether the relationship between $R_{AA}$ and $v_2$ is altered by subdiffusion and to quantify the potential effects on observables for both charm and beauty quarks. I believe that subdiffusion could significantly impact these observables. I plan to discuss these issues in detail in future articles.

\section{Acknowledgements} 

I acknowledge Dr. Santosh Kumar Das for his valuable suggestion, Aditi Tomar for her insightful discussions and inspirational motivation, and Mohammad Yousuf Jamal for his numerous informative contributions, all of which have collectively improved the content of this article. I acknowledge the Tata Institute of Fundamental Research, Mumbai, for a visit. Additionally, I acknowledge support from the SERB Fellowship Project Code No. RD/0122-SERBF30-001.

\section{Appendix : L1 NUMERICAL METHOD FOR SUbDIFFUSION PROCESS}
\label{AP}
Let $\{t_n = \frac{n}{N}T,;0 \leq n \leq N\}$ be a steps of time interval $ [0,T]$. And $^{C} D^\mu_{n}$ is the corresponding formula for the Caputo fractional derivative. For the case $0<\mu\leq 1$, the  L1 scheme is used. In this case, the first derivative $u^{(1)}$ is approximated using a linear interpolation formula, and the resulting numerical method applies only the values of $u$ at the previous two time points $u^{n-1}$ and $u^{n-2}$. 
\begin{widetext}
For $0<\mu\leq 1$, subdiffusion
\begin{align}
\nonumber ^{C} D^\mu_{0+}u(t_n) &= \frac{1}{\Gamma({1-\mu})}\int_0^{t_n} \frac{u^{(1)}(s)}{(t_n-s)^{\mu}}ds \\
\nonumber&= \frac{1}{\Gamma({1-\mu})}\sum_{j=1}^n\int_{t_{j-1}}^{t_j} \frac{u^{(1)}(s)}{(t_n-s)^{\mu}}ds \\ 
\nonumber&\approx \frac{1}{\Gamma({1-\mu})}\sum_{j=1}^n\frac{u(t_j)-u(t_{j-1})}{t_j -t_{j-1}}\int_{t_{j-1}}^{t_j} \frac{1}{(t_n-s)^{\mu}}ds \\
\nonumber& =\frac{1}{\Gamma({2-\mu})} \sum_{j=1}^n \frac{(t_n-t_{j-1})^{1-\mu}- (t_n -t_j)^{1-\mu}}{t_j -t_{j-1}} (u(t_j)-u(t_{j-1}) )\\
& = \sum_{j=0}^{n-1} a_j (u(t_{n-j})-u(t_{n-j-1})),
\end{align}
\end{widetext}
where the coefficients $a_j =\frac{(j+1)^{1-\mu}-j^{1-\mu}}{\Gamma({2-\mu}) \Delta t^{\mu}}$ for $1 \leq j \leq n-1$ in the scheme are determined by the difference formula for the first derivative and are used to account for the fractional order.

\begin{widetext}

\begin{align}\label{AP_sub}
^{C} D^\mu_{n}u(t_n)=\begin{cases}\displaystyle\frac{u^1-u^0}{\Delta t^{\mu}\Gamma(2-\mu)} \; &:\;n=1 \\
\displaystyle\frac{u^n-u^{n-1}}{\Delta t^{\mu}\Gamma(2-\mu)} +\sum_{j=1}^{n-1} a_j(u^{n-j}-u^{n-j-1}) \; &:\;n \geq 2.
\end{cases}
\end{align}
\end{widetext}
   



\bibliography{ref2}
\end{document}